\documentstyle[]{l-aa}  

\thesaurus{03      		 
              (02.18.7; 	 
	      11.01.2;		 
   	      11.03.3;		 
	      11.09.4)}		 

\begin{document}
 
\title{Ca depletion and the presence of dust in large scale
nebulosities in radiogalaxies (II)\thanks{Based on observations
carried out at the European Southern Observatory 
(ESO, La Silla, Chile)}}

\author{
M.Villar-Mart\'\i n
\inst{1,}
\inst{2}
\and
L.Binette
\inst{3}
}
 
\offprints{M. Villar-Mart\'\i n, Dept. of Physics, University of
Sheffield, Sheffield S3~7RH, UK}

\institute{$^{1}$\,ST-ECF, Karl-Schwarschild-Str 2, D-85748 Garching, Germany \\$^{2}$\,Dept. of Physics, University of
Sheffield, Sheffield S3~7RH, UK\\
$^{3}$\,European Southern Observatory, Casilla 19001, Santiago 19, 
Chile (LBinette@eso.org)}
 
\date{ }
 
\maketitle

\markright{Internal dust in Large Scale Nebulosities in RGs (II)}
 
\begin{abstract}

We investigate here the origin of the gas observed in extended
emission line regions surrounding AGNs. We use the technique of
calcium depletion as a test to prove or disprove the existence of dust
in such a gas in order to discriminate between two main theories: (1)
a cooling process from a hotter X-ray emitting phase surrounding the
galaxy, (2) merging or tidal interaction between two or more
components. We have obtained long slit spectroscopy of a sample of
objects representative of different galaxy types although our main
interest focus on radio galaxies.  The spectral range 
was set to 
always include
the [CaII]$\lambda\lambda$7291,7324 doublet.  The faintness or absence
of such lines is interpreted as due to the depletion of calcium onto
the dust grains and, therefore, is a proof of the existence of dust
mixed with the gas in the EELRs.
  
\end{abstract}

\section{Introduction}

	Extended emission surrounding AGNs has been detected in many
active galaxies.  Such structures have spatial scales that vary from
1kpc to hundreds of kpc.  Many of the extended emission line regions
(EELRs) are found in elliptical galaxies which are not gas rich
objects, thus: where does the gas come from? Does it come from stellar
evolution as a consequence of stellar winds and SN explosions? Or, on
the contrary, is the origin external, and the gas is accreted from a
companion galaxy rich in gas or from a hot halo surrounding the
galaxy? In Paper~I (Villar-Mart\'\i n \& Binette 1996) we made a
review of the observational evidences which supports or contradict these
theories, being an external origin easier to reconcile with the
observations.  The morphological similarities with the systems of
filaments surrounding galaxies near the centers of rich clusters
suggests that the origin of the EELR result from cooling of the hot
gaseous halos which surrounds the central galaxies. On the other hand,
morphological and kinematical studies of the gas often suggest the
existence of a collision or merger in the recent past.

	The existence or not of dust mixed with the emitting gas has
direct implications about the most plausible scenario for the
formation of these extended ionized structures. In the scheme of the
cooling flow theory, the hard intracluster radiation field would
prevent dust from forming. If the material consists of galactic
debris, the dust/gas ratio is expected to have a value appropriate to
the chemical composition of a normal galaxy. Our aim is to check if
the dust does exist in EELRs.

	Ferland (1993) has proposed that the absence of the forbidden
[CaII] lines $\lambda\lambda$7291,7324 (F1 and F2 hereafter) can be
used to infer the presence of dust mixed with the gas in the Narrow
Line Region (see also Kingdon, Ferland \& Feibelman, 1995). Calcium is very sensitive to dust. It suffers strong
depletion into the dust grains and therefore, its abundance in the
gaseous phase (responsible of the line emission) is lower than in a
dust free nebula. The line emission should be fainter that expected, or
even undetectable.

\begin{table*}[htb]
\centering
\caption{Observing Log}
\begin{tabular}{lllllll} \hline

Name & RA(1950) & Dec(1950) & z & Comment &  Exp time (s) &  PA \\ \hline
NGC1052 & 02 38 37.33 & -08 28 09 & 0.005 & Liner & 1800 & 270\degr \\ 
NGC6215 & 16 46 47.0 & -58 54 30 & 0.005 & Liner & 2700 & 240\degr  \\ 
NGC7552 & 23 13 25.0 & -42 51 24 & 0.005 & Liner & 1200 & 270\degr  \\ 
NGC7714 & 23 33 40.59 & 01 52 42 & 0.009 & Starburst &  2700 & 270\degr  \\ 
PKS1404-267 (nuc)& 14 04 38 & -26 46 51 &  0.021 & RG  & 2700 & 270\degr  \\ 
PKS1404-267 (5" S) & & &       &  & 2700 & 270\degr  \\
PKS2014-55  &  20 14 06 & -55 48 52 & 0.061 & RG & 3600 & 190\degr  \\ 
PKS2152-69 (nuc) & 21 52 58 & -69 55 40 & 0.028    & RG & 2700 & 270\degr  \\ 
PKS2152-69 (cloud) &  & &       &  & 3600 & 290\degr  \\
PKS2158-380 & 21 58 17 &  -38 00 51  & 0.033 &  RG & 2700 & 270\degr  \\ 
PKS2356-61 & 23 56 29 & -61 11 42 & 0.096 & RG & 3300 & 285\degr  \\ 
PKS2300-18 & 23 00 23 & -18 57 36 & 0.129 &    RG & 2700 & 240\degr  \\ 
2A 0335+096  &  03 35 52 & 09 48 10 & 0.035 & CF & 5400 & 147\degr  \\ 
A2029   &  15 08 30 & 05 57 00 & 0.077 & CF & 2700 & 270\degr  \\ 
A2597   &  23 22 42 & -12 23 00 & 0.085 & CF+RG & 2700 & 197\degr  \\ 
\hline
\end{tabular}

\end{table*}

	In Paper~I, we proved that such test remains valid (and is very
sensitive) under the conditions of the EELRs
studied here. We studied in detail all the most plausible {\it
alternative} mechanisms to that of internal dust for explaining the
absence of [CaII] lines. No acceptable alternative solution was found
and we concluded in favour of the validity of the method initially
proposed by Ferland (1993). The observational results and their
interpretation are presented in this paper. We have applied this test
to a sample of objects of different types: radiogalaxies with extended
emission line regions, cooling flows and starburst galaxies. Our main
interest concerns to the EELR in radio galaxies.

	We describe the observations, data reduction and analysis of
the spectra in section 2. In section 3 we present the results of the
comparison between measurements and model predictions. A detailed
analysis of the nuclear spectra is presented in section 4 and
conclusions compose section 5.

\section{Observations and data reduction}

The observations were carried out on the nights 21-23 August on
1993. All the spectra were obtained at La Silla Observatory, Chile,
with the 3.6 m telescope, using the EFOSC 1 spectrograph with a CCD
detector (TEK\#26) of 512 x 512 pixels$^2$ of 27 $\mu$m$^2$. The slit
width was 1.5". The grism used was, R150, with a dispersion of 120 \-
\AA\ /mm, a wavelength bin of 3.3 \AA\ /pixel and covering a spectral
range of $\sim$ 6870-8560 \AA\ . The observing conditions where
photometric. Table 1 gives a log of the observations.

\subsection{Basic data reduction}

The reduction of the data was done using standard methods provided in
IRAF. The spectra were bias subtracted and divided by a flat-field
frame (dome flat-field). Illumination corrections along the slit were
found to be negligible.

 In general, we obtained three frames for each object and each slit
position, allowing the direct removal of cosmic ray events. For a
given object, all the frames corresponding to the same slit position
were averaged together. In the cases where only two frames were
available, the cosmic rays were removed visually, replacing the
affected pixels by the mean of the surrounding region.  The spectra
were calibrated in wavelength using comparison spectra of an HeAr arc
taken at the beginning and end of each night, and additionally before
and after each object. The wavelength calibration was done very
carefully in order to later subtract the sky as accurately as
possible.  The IRAF routine ``background'' was used to substract the
contribution of the sky by interpolation of the background detected in
windows close to and on both sides of the object. Using this method, the
sky features were successfully substracted.

\begin{figure*}
\includegraphics{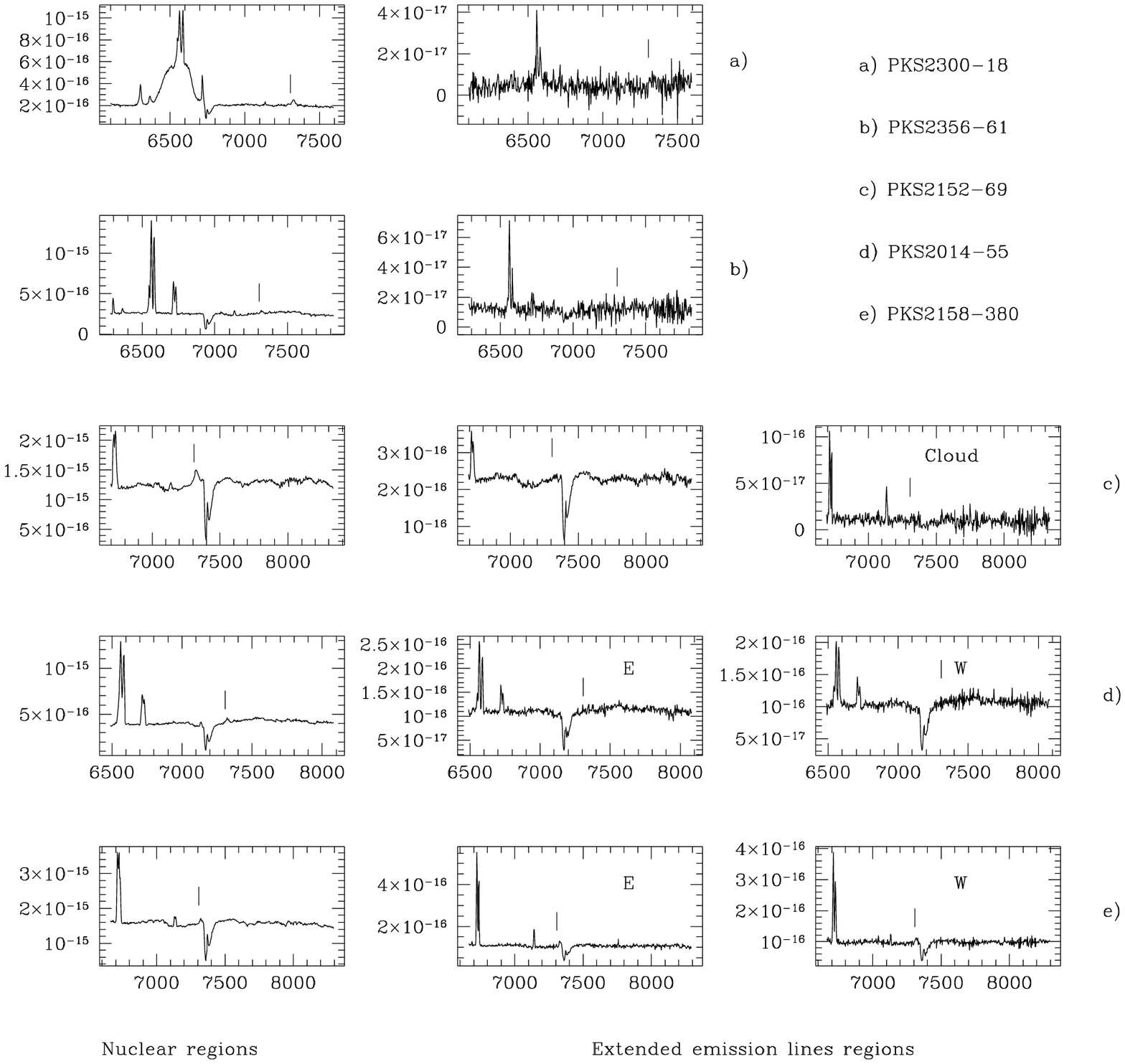}
\vspace{8in}

\caption[]{}{Spectra of observed Radiogalaxies, nuclear (first column)
and extended emission line regions (second and third columns). The
expected position of the F1 line is indicated with $\mid$ . Flux is
given in units of erg~ s$^{-1}$ cm$^{-2}$ \AA $^{-1}$.}

\end{figure*}

\begin{figure*}
\includegraphics{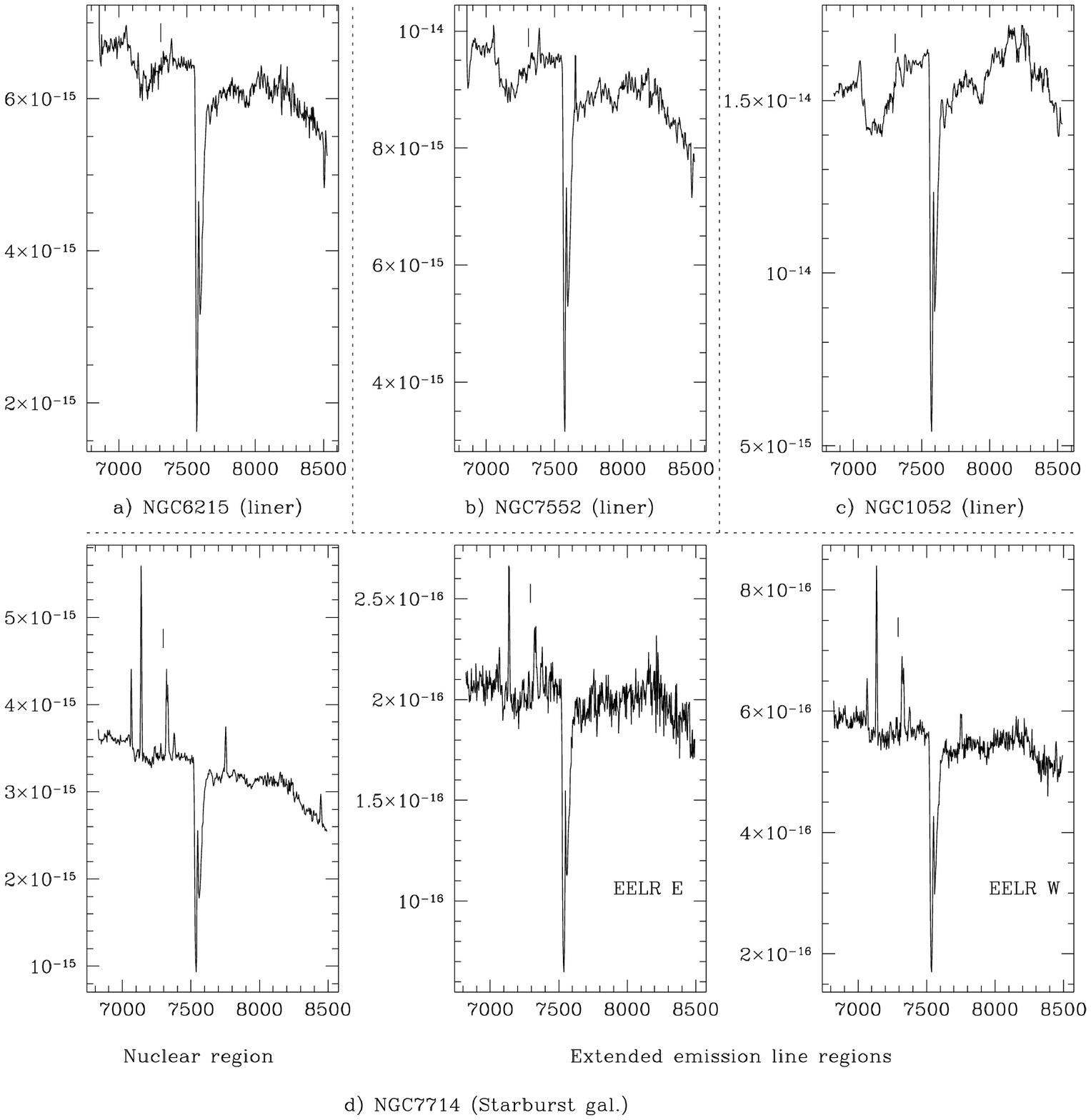}
\vspace{3.5in}

\caption[]{}{Spectra of observed Liners and Seyfert 2 galaxies. The
liner spectra have been extracted from the whole spatial extension of
the object. The expected position of the F1 line is indicated with
$\mid$ . Flux is given in units of erg~ s$^{-1}$ cm$^{-2}$ \AA $^{-1}$.}

\end{figure*}

\begin{figure*}
\includegraphics{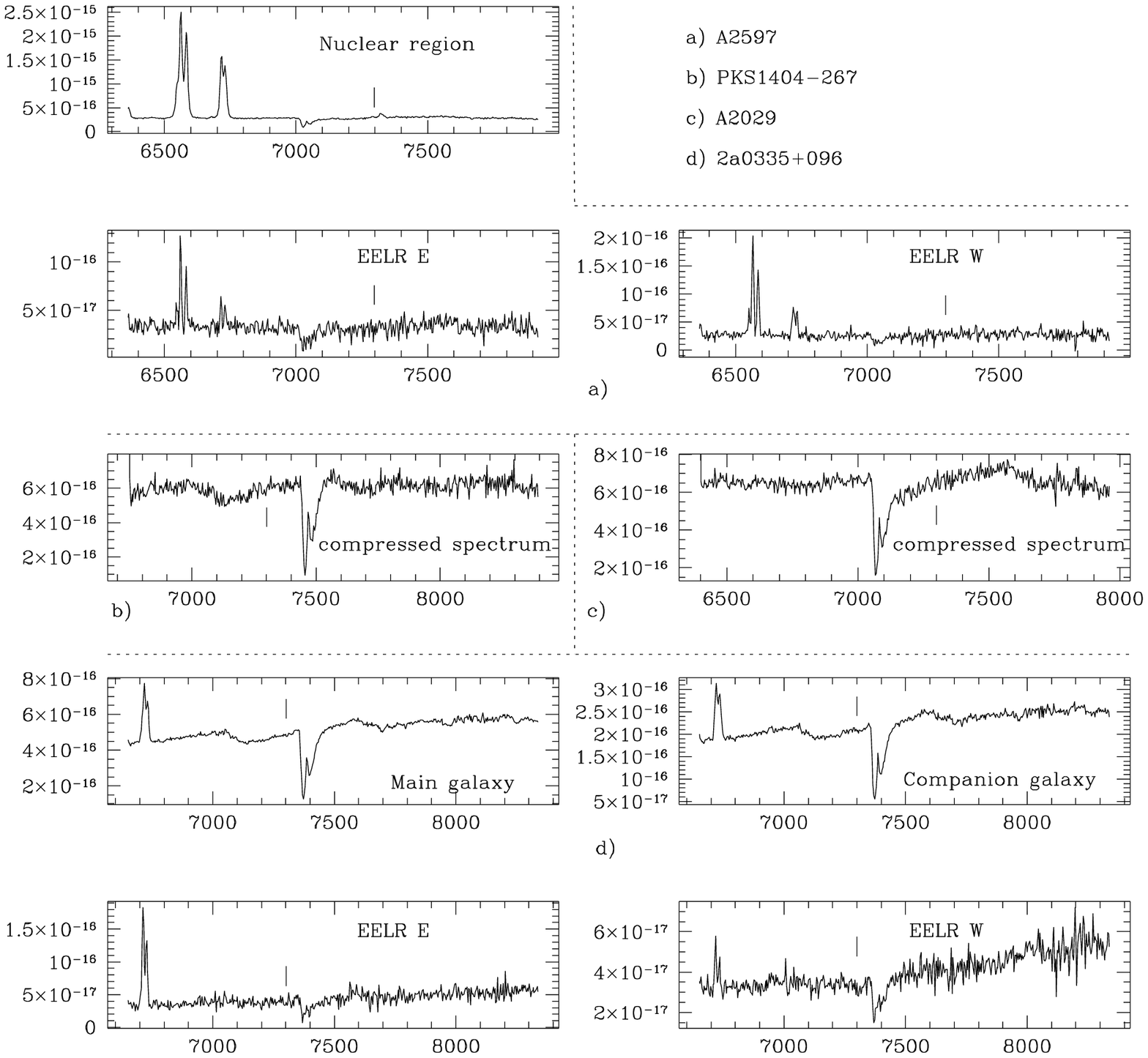}
\vspace{8in}

\caption[]{}{Spectra ot objects with properties associated with
cooling flows. The spectra of PKS1404-267 and A2029 have been
extracted from the whole spatial extension of the object.  The
expected position of the F1 line is indicated with $\mid$ . Flux is
given in units of erg~ s$^{-1}$ cm$^{-2}$ \AA $^{-1}$. }

\end{figure*}

\subsection{Atmospheric extinction}

	The spectra were corrected for atmospheric extinction with the
aid of mean extinction coefficients for La Silla. Molecular absorption
bands of O$_2$ (the A band at 7620 \AA\ and the H$_2$O bands, bands in
the region 7100-7450 \AA\ and 8100-8400 \AA\ ) were also evident in
the spectra. They were removed separately from all the observed
spectra. These absorption bands are composed of many closely spaced
absorption features, unresolved in our spectra. Many of the features
which comprise the A band are optically thick and therefore nearly
independent of zenith distance, but most of those in the H$_2$O bands
are not independent of time and zenith distance. In principle, the
best method to remove these lines is to observe standard stars as
close as possible in zenith distance and in time to that of the
program object in order to obtain the best correction
(c.f. Osterbrock, Shaw \& Veilleux 1990). But as these authors pointed
out, the star closest to the objects's observing time and zenith
distance is not always the one producing the best correcting spectrum.

	The procedure we used was the following: we observed three
different standard stars whose frames were reduced in the same way as
described before. After modeling the atmospheric bands with them, we
found that the best results were obtained with Feige 110, a dwarf with
no intrinsic absorption features in the spectral range observed. A
1-dimensional integrated spectrum of the standard star was obtained by
adding all the light along the spatial direction. A smooth fit to the
continuum was done and then the original spectrum was divided by the
fit. The result was a normalized spectrum which kept the features due
to atmospheric absorption. Due to the dependence on zenith distance
and time, we had to model the bands for each object, i.e. to construct
a specific restoring spectrum for each object, multiplying the
normalized spectrum by the appropriate factors.

\subsection{Flux calibration}

	The atmospherically corrected spectra of the standard stars
were used to obtain the flux calibration. For each night, we built a
mean response curve from the two standard stars observed that night.

\subsection{Template galaxy subtraction}

We checked that effects of stellar absorption features intrinsic to
the galaxy could be neglected: the calcium doublet is on top of the
raising part of a molecular band whose narrow line components are not
resolved. Its only effect is to change the slope of the continuum
under the doublet and we can easily correct for this effect, fitting a
smooth continuum of appropriate slope in such region. H$\alpha$, a
line which has been used as a reference in our prediction of the F1
line flux, can be underestimated due to the underlying H$\alpha$
absorption line. However, in our objects the H$\alpha$ emission is so
strong that the difference is less than 5\% of the emitted flux.  This
was estimated by comparing the EW of the emission line in our objects
with the EW of the absorption feature from a template elliptical
galaxy.

\subsection{Data analysis}

\subsubsection{Extraction of the spectra}

	For those objects for which no extended emission lines were
detected, we compressed the whole spatial extension in order to
extract a 1-dimensional spectrum. For those objects showing extended
emission lines, we separated it into two different components: the
nucleus, and the extended emission line spectrum. In order to achieve
this, we subtracted from the spatial profile of the brightest spectral
line, the spatial profile from the nearby continuum. In this way, we
are left with the spatial profile of the line emitting regions. In
some cases, emission blobs appear well detached from the nuclear
emission.  In other cases, however, the extended and nuclear emission
joined smoothly. In order to avoid contamination of the EELRs by the
unresolved nuclear spectrum, we fitted a Gaussian (PSF) to the nuclear
region. The residuals after subtracting this Gaussian were considered
to pertain to the EELR.  The resulting spectra are shown in Figs.~1, 2
and 3.

	It is possible that we may be venturing close enough to the
nucleus to sample higher densities, more typical of the classical
narrow line region (with densities higher than 10$^3$ cm$^{-3}$). To
verify this, when it was possible, we used the [SII] doublet to
estimate the density.  The values were always below 100 cm$^{-3}$,
which indicates that the spectra are dominated by low density gas.  In
any event, even if the NLR dominated the emission, calculations show
that at high densities the line F1 (in the dustfree case) if anything
becomes tronger than many other lines as a result of its very high
critical density and of the increase in electronic temperature. Our
conclusions about inferring dust or not are therefore not affected by
nuclear contamination. (Higher density models would in fact imply a
stronger calcium depletion).

\subsubsection{Line measurements}

	IRAF routines were used to measure the emission line
fluxes. For the blends, decomposition procedures were used, fitting
several Gaussians at the expected positions of the components. The
fluxes correspond to the values derived  from the sum of Gaussians which
better fitted the profile.

	The upper limits on [CaII] emission were measured directly
from the noise level in the data.

\begin{table*}
\centering
\footnotesize 

\caption{Emission line fluxes for the observed sample of galaxies  (in
units of erg~ s$^{-1}$ cm$^{-2}$ \AA $^{-1}$). The superscripts $^M$ and $^P$
indicate measured and predicted values respectively. The last two
columns show the predicted and measured values for the $F1$ line. For
those objects where the calculations were possible, the observed line
is clearly fainter than predicted by photoionization models. }

\vspace{0.5cm}

\begin{tabular}{llllclll} \hline
\centering

  Name & Spatial reg. & $ref~line$ & $Flux^{M}$  & ${\frac{F1}{ref~line}}^{P}$ & $Flux(F1)^{P}$ &  $Flux(F1)^{M}$  \\ \hline \hline
NGC1052 & Total  & None & --- & ---  & --- &   $\leq$~6.99e-16   \\ \hline
NGC6215 & Total & None & --- & --- & --- &   $\leq$~ 3.34e-16 \\ \hline
NGC7552 & Total  & None & --- & ---  & --- &   $\leq$~3.95e-16  \\ \hline
NGC7714 & Total  & [OII]$\lambda$7325 & 1.59e-14 & $\geq$1.00 & $\geq$ 1.59e-14 & $\leq$~6.84e-17   \\ \hline
PKS1404-267 & Nucleus & None & --- & ---  & ---  & ~ $\leq$~ 1.79e-16 \\ 
            & 5" South     & None   & ---  & ---  & ---    & ~ $\leq$~ 5.23e-17 \\ \hline
PKS2014-55 & Nucleus & [OII]$\lambda$7325 & 1.54e-15 & $\geq$1.00 & $\geq$1.54e-15 & $\leq$~3.34e-17 \\ 
        & EELR (E)      & H$\alpha$ &   1.47e-15 & $\sim 0.45$ & 6.62e-16 & $\leq$~ 2.60e-17   \\
        & EELR (W)      & H$\alpha$ &   1.06e-15 & $\sim 0.45$ & 4.77e-16 & $\leq$~  3.18e-17  \\ \hline
PKS2152-69 & Nucleus & [OII]$\lambda$7325 & 7.48e-15 & $\geq$1.00 & $\geq$7.48e-15 & 1.92e-16
  \\
         & EELR & [SII]$\lambda$6731 & 1.33e-15 & $\sim 0.40$ & 5.32e-16 & $\leq$~3.03e-17  \\
        &  Cloud & [ArIII]$\lambda$7136 & 3.86e-16  & $\sim 4.50$  & 1.74e-15 & $\leq$~2.24e-17   \\ \hline
PKS2158-380 & Nucleus &  [OII]$\lambda$7325 &   4.03e-15 & $\geq$1.00 & $\geq$4.03e-15 &
3.15e-16   \\
        & EELR (E) & [ArIII]$\lambda$7136 & 7.28e-16 & $\sim 5.0$ & 3.64e-15 &  $\leq$~2.55e-17  \\
        & EELR (W) & [ArIII]$\lambda$7136 & 3.00e-16 & $\sim 5.0$ & 1.5e-15 & $\leq$~5.00e-17 \\ \hline
PKS2356-61 & Nucleus & [OII]$\lambda$7325 &  9.39e-16 & $\geq$1.00 & $\geq$9.39e-16& 
2.78e-16 \\
        & EELR & H$\alpha$ & 5.36e-16 & $\sim 0.18$ &  9.6e-16 &  $\leq$~ 1.03e-17 \\ \hline
PKS2300-18 & Nucleus & [OII]$\lambda$7325 & 1.80e-15 & $\geq$1.00 &  $\geq$1.80e-15 & 4.37e-16 &    \\
        &  EELR & H$\alpha$ & 3.61e-16 & $\sim 0.40 $ & 1.44e-16 &   $\leq$~1.97e-17   \\ \hline
2A 0335+096  & cD galaxy & [SII]$\lambda$6725 & 7.46e-16  & $\sim 0.40$  & 2.98e-16 & $\leq$~ 2.72e-17 \\ 
        & EELR (E) & [SII]$\lambda$6725 & 2.24e-15 & $\sim 0.40$   & 8.96e-16 & $\leq$~1.29e-17  \\
        & EELR (W) & [SII]$\lambda$6725 & 4.22e-16 & $\sim 0.40$  & 1.68e-16 & $\leq$~1.40e-17   \\
        & companion  & [SII]$\lambda$6725 & 7.89e-15 & $\sim 0.40$  & 3.16e-15 & $\leq$~ 1.69e-17  \\ \hline
A2029   & Total  & None & --- & --- & ---  & $\leq$~1.43e-16  \\ \hline
A2597   & Nucleus & [OII]$\lambda$7325  & 1.79e-15 & $\geq$1.00 &  $\geq$1.79e-15  & 5.14e-16  \\ 
        & EELR & H$\alpha$ & 2.32e-15 & $\sim 0.50$&  1.16e-15 &  $\leq$~5.86e-17 \\ \hline
\hline

\end{tabular}
\end{table*}

\section{Observed and predicted F1 fluxes}

\subsection{Prediction of F1 fluxes}

	The second component of the [CaII] doublet (F2) is the weakest
and is blended with the [OII]$\lambda\lambda$7320,7330 multiplet. The
first component is the strongest and lies well aside, some 30 \AA\
shortward of [OII]. It is straightforward to isolate it given a
reasonable spectral resolution. Therefore, for simplicity, we have
based our study on the measurement of this line. The observed values
are compared with the flux that photoionization models
predict. MAPPINGS (Binette et al. 1993a,b) is the multipurpose
photoionization-shock code which we used for the modeling. It considers
the effects of dust mixed with the ionized gas: extinction of the
ionizing continuum and of the emission lines, scattering by the dust,
heating by dust photoionization and depletion of heavy elements. The
appropriate input parameters for the models are justified in Paper~I.

	The predicted value of the F1 flux has been computed in the
following way: we used as reference line ($ref~line$), the strongest
and/or the easiest line to measure, like H$\alpha$,
[SII]$\lambda\lambda$6716,6731, the [OII]$\lambda\lambda$7320,7330
blend and/or [ArIII]$\lambda$7136.  The appropriate photoionization
model allows the prediction of $R=\frac{\rm{F1}}{ref.~line}$, from
which we can easily deduct the F1 flux:

{\it Predicted} F1 {\it flux =} {\it R
$\times$ Measured flux in the reference line.}

	In Table 2, we show for each object and for each spatial
region the reference lines used, the measured flux, the predicted
$R=\frac{\rm{F1}}{ref.~line}$ ratio, the measured F1 flux and its
theoretical value. For the nuclear regions, we always base our
predictions on the blend [OII] $\lambda\lambda$7320,7330 as reference
line.  The density in the nuclear narrow line region spans a wide
range up to 10$^6$ cm$^{-3}$. The critical density of the [CaII]
forbidden lines is very high, and the flux of the lines under the
nuclear conditions are predicted to be far stronger than the [OII]
lines (Ferland 1993). The low critical density [OII] line can be
de-excited as a result of the high nuclear density so that if [CaII]
remains much smaller than [OII], it is actually a more stringent test
since high densities would have actually helped to increase the
[CaII]/[OII] ratio.  To measure the flux in the [OII] blend, we assume
that there is a negligible contribution from the F2 component. This
assumption is reasonable: it is fainter that F1 and, in any event, it 
is near
or under the detection limit in all cases. We can therefore assume
that the [OII] multiplet is not contaminated by F2.

\subsection{Comparison with observations}

	The comparison between the measured and predicted F1 flux
values shown in table 2 demonstrates that, whenever the calculations
were possible (sometimes, however, there was not any reference line
available), it is always fainter (non detected in most cases) than
expected. This result is common to all the regions considered here:
nuclear, EELRs and cooling flow filaments. It demonstrates that
calcium is depleted and, therefore, {\it the gas is mixed with dust},
both in the nuclear as well as in the extended emitting gas in radio
galaxies and cooling flow filaments.
 
Donahue \& Voit applied this same test to the emission line nebulae in
cluster cooling flows (1993). They interpreted the observed lack of
[CaII] emission as depletion within the ionized filaments of calcium
onto dust grains.  Although for quite a different sample of objects,
our work support these results, namely that the {\it ionized gas}
which emits the low excitation lines observed in nuclear regions of
AGNs, in large scale ionized gas in RGs, and in the filaments in
objects classified as `cooling flows' must all contain dust.

The most important result is the confirmation that {\it dust exists
mixed with the ISM in low $z$ radio galaxies}.

\section{Discussion}

\subsection{Implications on the origin of the gas}

As explained in Paper~I, there are some pieces of evidence indicative
of dust {\it in the extended ionized gas} in active galaxies, mainly
derived from polarization measurements, which show the existence of
scattered nuclear light over large spatial scales, although it is
difficult to discriminate between dust and electrons as the scattering
agent.  Our results confirm that dust exists {\it mixed with the ISM
in low redshift radio galaxies}.

As mentioned before, this produces discrepancies with the traditional
cooling flow theory, which would now be required to explain the
formation of dust in a shorter time than the cooling time! According
to this theory (see Fabian 1994 for a review), the galaxies, groups
and clusters of galaxies were formed out of gas that collapsed
gravitationally. During this process, gravitational energy was
released which heated the clouds and a hot X-ray atmosphere
resulted. Inhomogeneities in the gas cause matter from the hot
atmosphere to cool down and fall towards the center of the galaxy or
cluster. In this picture, the resulting filaments would eventually
become visible at optical wavelengths and would emit strong lines. For
massive galaxies and cluster, the cooling process would have been
slower than for normal galaxies and the hot atmosphere would still
exist, with typical temperatures of several million K, and be emitting
strongly in the X-ray band.

If this is actually the origin of the gas in the EELRs (and cooling
flow filaments), it should clearly be devoid of dust.  Any dust
introduced in the hot intracluster medium would be sputtered and
rapidly destroyed (time scale of the order of 10$^7$ years) (Draine \&
Salpeter 1979), much before the filaments cooled down and became
visible.  Is there a way to introduce dust in the filaments during the
cooling process? The common place where dust is formed is stellar
atmospheres. Do stars exist in the accreted gas?  Such gas must be
deposited in some form and it has been generally thought that the more
plausible fate for most of the gas is the formation of new stars
(e.g., O'Connell \& McNamara 1989; Fabian, Nulsen \& Canizares 1991).
Although there are some indications of star formation in cooling flow
galaxies, the latter seems to be taking place in the inner parts (over
the central few kpc) (Cardiel, Gorgas \& Arag\'on-Salamanca 1995;
Fabian 1994).  There is a lack of evidence of star formation (or any
kind of stars) at large distances, out of the main body of the
galaxies, where the EELRs filaments are still visible.

In high z RGs, there is an extended blue continuum aligned with the
radio axis which has often been interpreted as due to young
stars. However, as I mentioned already before, further results (e.g.,
Tadhunter, Fosbury \& di Serego 1988; Cimatti et~al. 1993) proved that
this continuum is polarized and that the contribution of a hidden
quasar continuum scattered by dust and/or electrons in the ISM
constitutes very likely an important component (in fact, this extended
polarized UV continuum has been used as a proof of the existence of
dust in the EELRs of high z radio galaxies). This effect has also been
detected in some radio galaxies at low $z$ (Cimatti \& di Serego
Alighieri S. 1995).  Altough young stars may exist, we don't have a
clear idea about their overall contribution.

	There is another mechanism which can form dust: if most of the
cooled gas from a flow does not form stars with normal IMF, maybe it
remains as cold clouds or as low-mass stars.  There are
evidences of X-ray absorption (White et al. 1991; Mushotzky 1992;
Allen et al. 1993) in cooling flows in cluster of galaxies.  The
absorbing material could be in the form of cold gas embedded in the
hot intracluster medium of the cooling flow (White et al. 1991;
Daines, Fabian \& Thomas 1994). This cold gas, very slightly
photoionized by X-rays from the surrounding hot corona, can become
molecular (Ferland, Fabian \& Johnstone, 1994). Fabian, Nulsen \&
Canizares (1994) propose that the conditions suitable for dust to form
may occur in this cool gas, through the condensation of gaseous
particles. The authors propose that the gas which is cooling towards the center
could be a mixture of both the molecular gas and the very hot gas and,
therefore, could contain the existing dust. However, even if such a scheme was
possible, the dust grains will not remove the calcium which already existed in the hot gas; the temperature is too high for the condensation of calcium 
onto the dust grains. Therefore, we should observe the CaII lines from the hot gas when it cools down, even if it contains dust.

	On the other hand, the interaction scenario (gas accreted from
outside the galaxy, as a result of recent tidal interactions or
mergers between two or more components) predicts the existence of dust
mixed with the gas, the one already existing in the interacting
objects. Some morphological evidences and theoretical ideas (see
Paper~I) support this scene.  Heckman et al. (1986) showed that a
large fraction of powerful radio galaxies have morphological features
(shells, tails, loops, etc) similar to those produced in numerical
simulations of galaxy interactions (e.g., Toomre and Toomre 1972;
Quinn 1984). Kinematic measurements show that the radio galaxy EELR
generally have a high specific angular momentum which is difficult to
reconcile with the cooling flow picture (Tadhunter, Fosbury \& Quinn
1989).

	Therefore, if there is a common origin for the EELRs of all
radio galaxies, our results suggest that it is mergers or tidal
interactions. However, we don't exclude the possibility of an origin
which is {\it not} universal, that is, which may differ from object to
object.

\section{Conclusions}

	We have confirmed that the gas in extended emission line
regions in radio galaxies at low $z$ is mixed with dust. 

	Our results support the existence of dust mixed with the gas
in the Narrow Line Region and in the cooling flow filaments.
	
	If there is an universal origin for the EELRs, the existence
of internal dust favours mergers or tidal interactions as the most
plausible scenario.

\begin{acknowledgements} MV-M acknowledges support from the Deustche
Forschungsgemeinschaft; also thanks to A. Caulet, R.A.E. Fosbury,
J. van Loon, R. Pelletier and I. P\'erez-Fourn\'on for useful
discussions.

\end{acknowledgements}

\end{document}